\documentclass[a4paper,11pt]{article}
\usepackage{pos}
\usepackage{wasysym}
\usepackage{siunitx}

\newlength{\bibitemsep}
\newlength{\bibparskip}
\let\oldthebibliography\thebibliography
\renewcommand\thebibliography[1]{%
  \oldthebibliography{#1}%
  \setlength{\parskip}{\bibitemsep}%
  \setlength{\itemsep}{\bibparskip}%
}

\title{Likelihood reconstruction of radio signals of neutrinos and cosmic rays}

\author*[a]{Martin Langgård Ravn}
\author[a]{Chistrian Glaser}
\author[a]{Thorsten Glüsenkamp}
\author[a]{Alan Coleman}

\affiliation[a]{Dept. of Physics and Astronomy, Uppsala University, Box 516, S-75120 Uppsala, Sweden}

\emailAdd{martin.ravn@physics.uu.se}

\abstract{Ultra-high-energy neutrinos and cosmic rays are excellent probes of astroparticle physics phenomena. For astroparticle physics analyses, robust and accurate reconstruction of signal parameters like arrival direction and energy is essential. Current reconstruction methods ignore bin-to-bin noise correlations, which limits reconstruction resolution and so far has prevented calculations of event-by-event uncertainties. In this work, we present a likelihood description of neutrino or cosmic-ray signals in a radio detector with correlated noise, as present in all neutrino and cosmic-ray radio detectors. We demonstrate with a toy-model reconstruction that signal parameters such as energy and direction, including event-by-event uncertainties with correct coverage, can be obtained. Additionally, by correctly accounting for correlations, the likelihood description constrains the best-fit parameters better than alternative methods and thus improves experimental reconstruction capabilities.}

\FullConference{
  10th International Workshop on Acoustic and Radio EeV Neutrino Detection Activities\\
  11 - 14 June, 2024\\
  The Kavli Institute for Cosmological Physics, Chicago, IL, USA
}

\begin{document}
\maketitle

\section{Introduction}
In this contribution, we present a probabilistic description of voltage time-traces recorded in radio detectors with band-limited noise and a deterministic neutrino or cosmic-ray signal. The probabilistic description can be viewed as the likelihood for a signal given the measured traces and is thus the optimal objective function for reconstruction. We demonstrate with a toy-model that, when using the likelihood description, the signal parameters can be reconstructed with uncertainties, the uncertainties give correct coverage, and the reconstruction resolution improves compared to previous methods.

In recent years, radio detectors have proven to be powerful tools for probing ultra-high-energy astroparticle phenomena. The radio detection technique has been successfully employed in cosmic-ray air shower observatories, such as the Pierre Auger Observatory and LOFAR, and has yielded accurate estimation of arrival direction, shower maximum, and energy with precision competitive to other established methods \cite{Schroder:2016hrv}. In the field of neutrino astronomy, radio detectors have emerged as the most promising approach to detect the yet undiscovered ultra-high-energy neutrino flux at PeV energies and above \cite{Barwick:2022vqt}. Several prototype in-ice radio arrays, such as ARIANNA and ARA, along with balloon experiments, like ANITA, have demonstrated the viability of this technique. Furthermore, the construction of larger in-ice arrays, such as RNO-G, is ongoing.

In any astroparticle physics analysis involving radio-detected signals, robust and accurate estimation of the signal parameters is crucial. For cosmic-ray and neutrino signals, the parameters of interest are the energy of the primary particle and its arrival direction. Additionally, for cosmic-ray studies, the shower maximum is an important proxy for the composition.

Several successful methods for reconstructing neutrino or cosmic-ray signals in radio detectors have been presented in recent years. These are either based on summary statistics calculated for each antenna observing the signal (e.g. signal arrival time, maximum amplitude, and fluence), or they use the full time-trace measured by each antenna. In general, the time-traces hold all available information and should result in optimal reconstruction, however, since antenna responses and other detector effects can be challenging to model and measure, methods using summary statistics sometimes result in a more stable reconstruction.

The current method for in-ice radio detectors is the \emph{forward folding technique} \cite{Glaser:2019rxw}, in which the radio emission of the neutrino or cosmic-ray interaction is calculated, propagated through the given medium, and folded through the antenna and hardware response to obtain the signal as it would appear in the recorded time-trace. The predicted signal is then compared to a measured time-trace, where bin-by-bin the chi-square between the predicted and measured voltage is calculated and summed over all bins and antennas:
\begin{equation} \label{eq_chi2}
    \chi^2 = \sum_{\text{i} = 1}^{n_{\text{ant}}} \sum_{\text{k=0}}^{n_t-1} \frac{[V_{i,k} - \mu_{i,k}(\theta)]^2}{\sigma_k^2},
\end{equation}
where $V_{i,k}$ is the measured voltage in the $i^{\rm th}$ antenna and $k^{\rm th}$ time bin, $\mu_{i,k}(\theta)$ is the predicted signal parameterized by $\theta$, $\sigma_k$ is the root mean square (RMS) of the noise in the $k^{\rm th}$ antenna, $n_{\text{ant}}$ is the number of antennas, and $n_{t}$ is the number of time bins. This $\chi^2$ is then used as an objective function to minimize in a reconstruction of signal parameters, which has been shown to work well for reconstruction of electrical fields from cosmic rays and neutrino signals \cite{Glaser:2019rxw, Plaisier:2023cxz, Arianna:2021lnr}. While references \cite{Glaser:2019rxw, Plaisier:2023cxz} are simulation studies, reference \cite{Arianna:2021lnr} used the \emph{forward folding technique} to reconstruct cosmic rays measured by the ARIANNA detector and demonstrates the applicability of the method under experimental conditions.

However, the $\chi^2$ expressed in Equation~\eqref{eq_chi2} can not be used to estimate uncertainties on the reconstructed parameters on an event-by-event basis. The reason for this is that the noise present in radio detectors is not white noise, but can, in general, be characterized as band-limited, which results in correlations between neighboring bins in the time domain. Since the correlations are not taken into account in the $\chi^2$, it is not $\chi^2$-distributed and can thus not be used to estimate uncertainties. Instead, the correct probabilistic description of band-limited noise plus a deterministic signal, which takes the correlations in the noise into account, is needed to calculate the likelihood for a signal. We present such a model in this work.

\section{Probabilistic noise model} \label{sec_noise_model}
\begin{figure}[h]
	%\makebox[\linewidth][c]{
	\centering
	\includegraphics[width=1\textwidth]{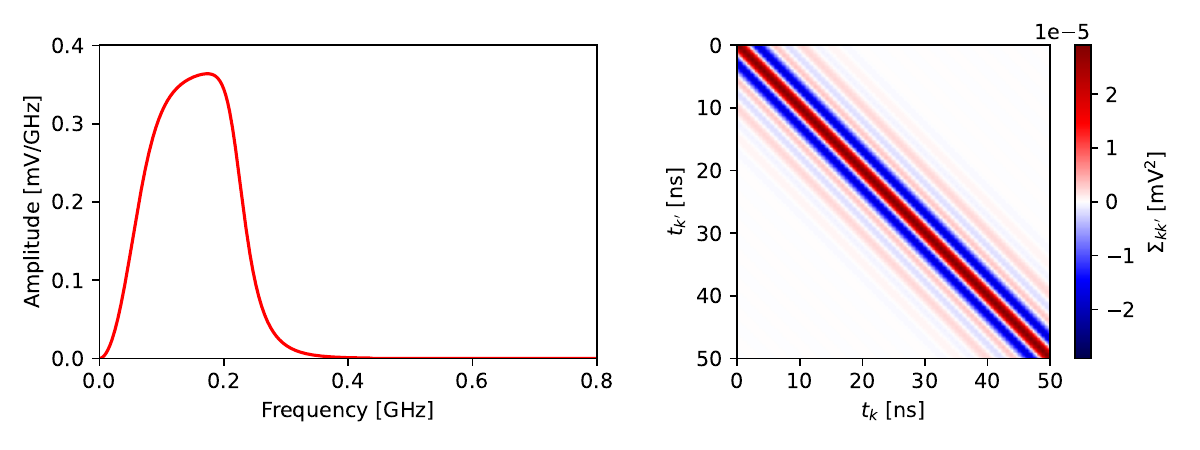}
	\caption{Left: Band-limited frequency spectrum, which is typical for radio detectors. Frequency amplitudes below $1\permil$ of the maximum height of the spectrum are set to 0 to avoid numerical instabilities. Right: Covariance matrix corresponding to the spectrum represented as an image (zoom in on the first 50 ns).}
	\label{fig_covariance_matrix}
\end{figure}

We model the band-limited noise plus a deterministic signal in a time-trace as a multivariate normal distribution with dimensionality equal to the length of the trace, $n_t$. The probability density of measuring any trace $\boldsymbol{x} = (V_{0}, ..., V_{n_t-1})$ is then:
\begin{equation} \label{eq_multivariate_normal}
	p\big(\boldsymbol{x};\boldsymbol{\mu}(\theta),\boldsymbol{\Sigma}\big) = \frac{1}{\sqrt{(2\pi)^{n_t} |{\boldsymbol{\Sigma}}| }} \exp \bigg(-\frac{1}{2}\big(\boldsymbol{x}-\boldsymbol{\mu}(\theta)\big)^\text{T} \boldsymbol{\Sigma}^{-1} \big(\boldsymbol{x}-\boldsymbol{\mu}(\theta)\big) \bigg),
\end{equation}
where $\boldsymbol{\mu}(\theta) = (\mu_{0}(\theta), ..., \mu_{n_t-1}(\theta))$ is the signal (zeros if no signal is present) parameterized by $\theta$, $\boldsymbol{\Sigma}$ is the covariance matrix of the noise, and $|{\boldsymbol{\Sigma}}|$ is its determinant. The covariance matrix can be estimated from many traces consisting purely of noise via the estimate of the autocovariance function:
\begin{equation} \label{eq_covariance_empirical}
    \mathrm{Cov}(t_k,t_{k'}) = \frac{1}{N}\sum_{n=1}^{N} (x_{n,k}-\bar{x}_k)(x_{n,k'}-\bar{x}_{k'}),
\end{equation}
where $N$ is the number of traces and $\bar{x}_{k}$ is the mean of the $k^{\rm th}$ time bin, which is zero for traces of pure noise. By assuming the covariance matrix is a symmetric circulant matrix, the elements only depend on $\Delta t_m = t_{k'} - t_k$, where $m = k' - k$, and we can average over elements with the same $\Delta t_m$:
\begin{equation} \label{eq_covariance_averaged}
    \langle \mathrm{Cov}(\Delta t_{m}) \rangle = \frac{1}{n_t} \sum_{k=0}^{n_t-1} \sum_{k'=0}^{n_t-1} \delta_{k,(k'-m)} \mathrm{Cov}(t_k,t_{k'}) .
\end{equation}
where $\delta$ is the Kronecker delta. The elements of the covariance matrix are then:
\begin{equation} \label{eq_covariance_matrix_elements}
    \boldsymbol{\Sigma}_{k k'} = \langle \mathrm{Cov}(\Delta t_{k' - k})\rangle .
\end{equation}
The circulant structure of the covariance matrix reflects that the noise correlations are invariant under time translation and are periodic. Although the periodic assumption is not physically accurate, since it implies correlations between the first and last bins of a trace, it has minimal impact on this work as the signals considered can be assumed not to be close to the edges of the traces. Alternatively, a Toeplitz structure can be employed, which does not impose the periodic constraint. However, the circulant structure is more numerically stable and simplifies the implementation, and is thus used in the following. An example of a typical frequency spectrum for radio detectors and the corresponding covariance matrix estimated from 10000 realizations of noise are shown in Figure~\ref{fig_covariance_matrix}.

We note that the multivariate normal distribution (Equation~\eqref{eq_multivariate_normal}) depends on the inverse of the covariance matrix, which can only be calculated if it is full rank. The covariance matrix is, however, not full rank when any frequency amplitudes are exactly zero. In this case, the multivariate normal distribution is calculated by replacing the inverse with the Moore-Penrose pseudo inverse $\boldsymbol{\Sigma}^{-1} \rightarrow \boldsymbol{\Sigma}^{+}$, the determinant with the pseudo-determinant $|\boldsymbol{\Sigma}| \rightarrow |\boldsymbol{\Sigma}|_{+}$, and $n_t$ with two times the number of non-zero frequency amplitudes $n_t \rightarrow n_{\text{dof}}$, which is the number of degrees of freedom.

\section{Likelihood reconstruction}
The multivariate normal probability of Equation~\eqref{eq_multivariate_normal} can be viewed as the likelihood for a signal given a trace. When the signal is present in many antennas, the full likelihood for the signal is the product of the likelihood for each antenna in the detector. The minus two log-likelihood is then:
\begin{equation} \label{eq_minus_two_delta_llh}
	\begin{aligned}
		-2\ln{\mathcal{L}(\boldsymbol{\mu}(\theta);\boldsymbol{x},\boldsymbol{\Sigma})} &= \sum_{i=1}^{n_{\text{ant}}} \big(\boldsymbol{x}_i-\boldsymbol{\mu}_i(\theta)\big)^\text{T} \boldsymbol{\Sigma}_i^{-1}  \big(\boldsymbol{x}_i-\boldsymbol{\mu}_i(\theta)\big) + const.
	\end{aligned}
\end{equation}
where $\boldsymbol{x}$, $\boldsymbol{\mu}_i (\theta)$, and $\boldsymbol{\Sigma}_i$ are the trace, signal, and covariance matrix of the $i^{\rm th}$ antenna, and the constants can be ignored since, in general, only delta log-likelihoods are of interest, i.e., likelihood ratios. To demonstrate that the likelihood description can be used to reconstruct signals and estimate uncertainties on the reconstructed parameters, we perform a toy-model reconstruction of a simplified neutrino signal in a toy-model in-ice radio detector. The neutrino signal calculation and detector simulation are performed using NuRadioMC \cite{Glaser:2019cws} and NuRadioReco \cite{Glaser:2019rxw}.

The toy-model detector layout is a typical ``deep'' station layout similar to the design of the ARA detector at the South Pole \cite{Allison:2011wk} and RNO-G in Greenland \cite{RNO-G:2020rmc}, and similar to the deep detector component foreseen for IceCube-Gen2 \cite{Gen2TDR}. It consists of 12 antennas in total, located in three holes in the ice. Each string has both a vertically polarized (VPol) and a horizontally polarized (HPol) antenna at 80\,m depth and the same at 100\,m depth. The coordinates of the strings in the horizontal plane ($x$,$y$) are (0\,m, 0\,m), (20\,m, 0\,m), and (0\,m, 20\,m) which form a right isosceles triangle. %This station layout enables reconstruction of the neutrino signal parameters where the neutrino signal is present in several antennas.

The Askaryan radio emission from the neutrino interaction is calculated using the \textit{Alvarez2009} \cite{Alvarez-Muniz:2010hbb} parameterization as implemented in NuRadioMC, assuming a hadronic shower. The most computationally intensive part of the signal calculation is the propagation of the radio signal through the ice. For ease of computation, we thus neglect the refraction of radio waves in non-uniform media and assume the signal travels in a straight line from the interaction vertex to the detector. This simplification does not affect the conclusions of this work. After the signal is propagated through the ice, it is folded through the antenna response, and noise is added to the resulting trace. The noise is generated with the frequency spectrum shown in Figure \ref{fig_covariance_matrix}, which is typical for radio detectors. We set the noise temperature to \SI{300}{K}. The traces are $n_t = 1024$ samples long and have a sampling rate of \SI{1.6}{GHz}.

\begin{figure}[h]
	%\makebox[\linewidth][c]{
	\centering
	\includegraphics[width=1\textwidth]{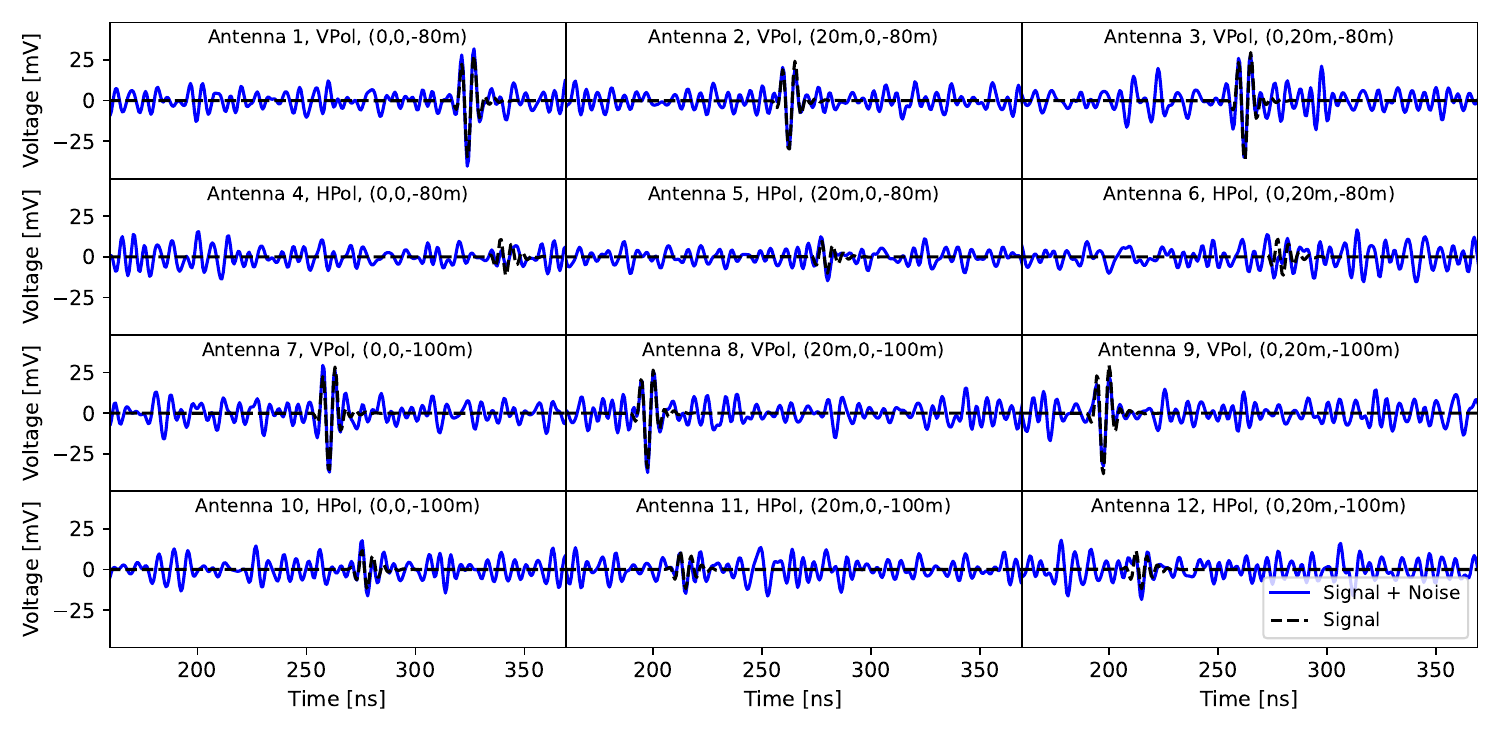}
	\caption{Traces for the 12 antennas of the toy-model in-ice radio detector with a neutrino signal (Event 3 in Table \ref{tab_events}) and band-limited noise (left panel in Figure \ref{fig_covariance_matrix}).}
	\label{fig_signal}
\end{figure}

The resulting neutrino signal observed in the detector depends on seven parameters: the shower energy, $E_\text{shower}$, the neutrino arrival direction zenith and azimuth angles, $\theta_\nu$ and $\phi_\nu$, the interaction vertex position in spherical coordinates, $r_\text{vertex}$, $\theta_\text{vertex}$ and $\phi_\text{vertex}$, and the signal arrival time, $t_0$. These are the parameters we aim to estimate in a reconstruction. In this work, the results for three representative example events are shown, labeled event 1, 2, and 3. The Monte Carlo truth values of the parameters for the three events are shown in Table~\ref{tab_events}. The traces in all 12 antennas for Event 3 with one realization of noise is shown in Figure~\ref{fig_signal}.

\begin{table}[!htb]
      \centering
        \begin{tabular}{|c||c|c|c|c|c|c|c|} 
            \hline
            Event & $E_\text{shower}$ & $\theta_\nu$ & $\phi_\nu$ & $r_\text{vertex}$ & $\theta_\text{vertex}$ & $\phi_\text{vertex}$ & $t_0$ \\ \hline \hline
            1 & 100\,PeV & $70^{\circ}$ & $25^{\circ}$ & 800\,m & $135^{\circ}$ & $30^{\circ}$ & 200\,ns \\ \hline
            2 & 100\,PeV & $70^{\circ}$ & $45^{\circ}$ & 800\,m & $100^{\circ}$ & $100^{\circ}$ & 200\,ns \\ \hline
            3 & 100\,PeV & $90^{\circ}$ & $1^{\circ}$ & 700\,m & $135^{\circ}$ & $45^{\circ}$ & 200\,ns \\ \hline
        \end{tabular}
    \caption{True parameter values for the three example events shown in this work.}
     \label{tab_events}
\end{table}
\vspace{-0.5cm}

The reconstruction is performed by minimizing the $-2\ln{\mathcal{L}(\boldsymbol{\mu}(\theta);\boldsymbol{x},\boldsymbol{\Sigma})}$ of Equation~\eqref{eq_minus_two_delta_llh} with respect to the neutrino signal parameters $\theta = (E_\text{shower}, \theta_\nu, \phi_\nu, r_\text{vertex}, \theta_\text{vertex}, \phi_\text{vertex}, t_0)$ using the iminuit \cite{iminuit} MIGRAD algorithm. To investigate the ideal case where the global minimum of the likelihood can be found, the parameters are initialized at the true values in the minimization process.

\begin{figure}[h]
	%\makebox[\linewidth][c]{
	\centering
    \includegraphics[width=1\textwidth]{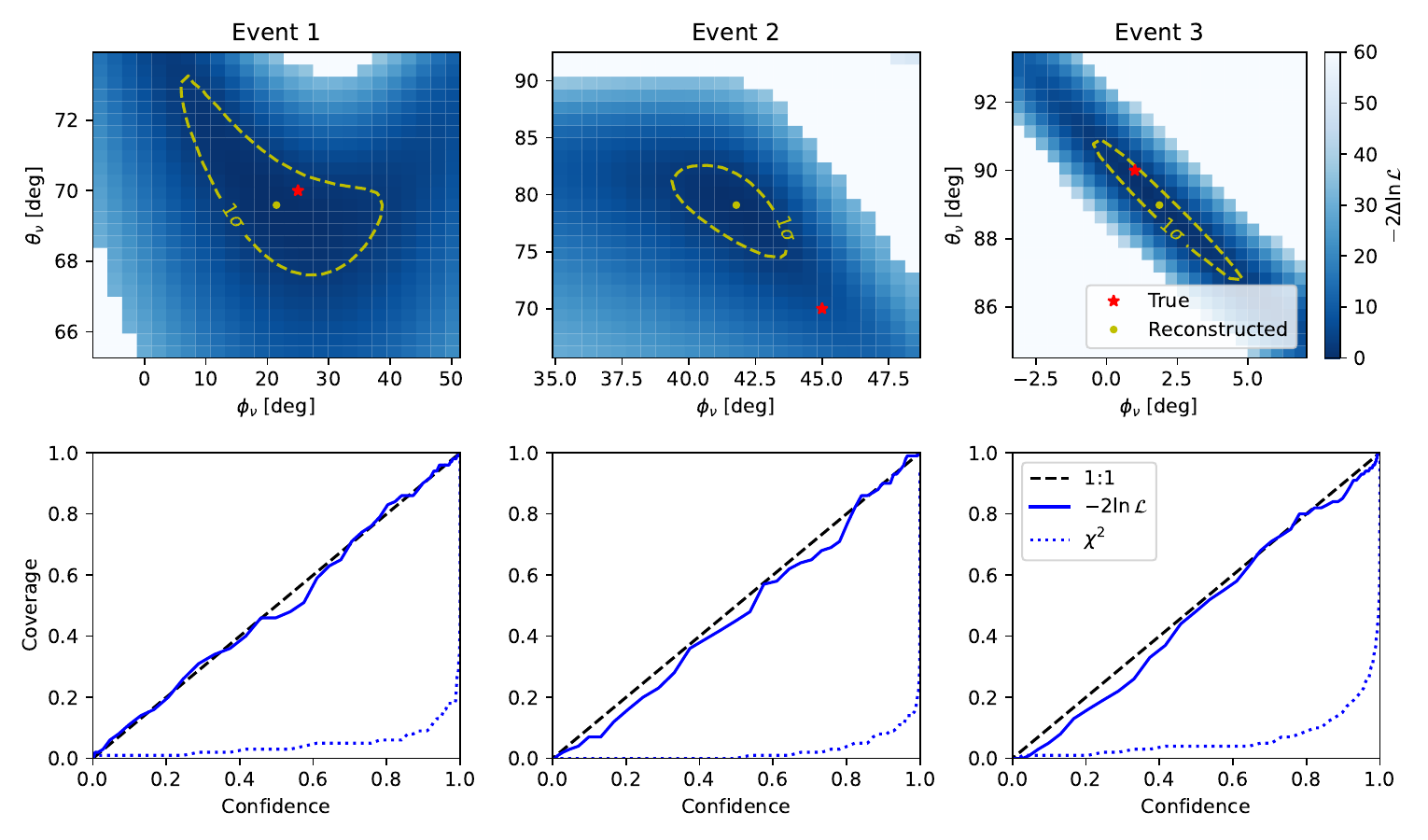}
	\caption{Top: Profile likelihood scans for the three example events listed in Table \ref{tab_events} with one realization of noise each along with the reconstructed arrival direction, $1\sigma$ uncertainty contours, and the true arrival direction. Bottom: Coverage for 100 trials of the same neutrino signals with different realizations of noise for the three example events using the $-2\ln{\mathcal{L}}$ expressed in Equation~\eqref{eq_minus_two_delta_llh} presented in this work along with the coverage using the $\chi^2$ with no correlation taken into account shown in Equation~\eqref{eq_chi2}.}
	\label{fig_profile_llh}
\end{figure}

To estimate uncertainties on a subset of the reconstructed parameters, we perform a profile likelihood scan. In many physics analyses, the primary parameters of interest are the neutrino energy or the arrival direction, while the remaining parameters can be treated as nuisance parameters and profiled over. In this study, we focus specifically on the neutrino arrival direction. The profile likelihood scan procedure is as follows: First, the reconstruction is performed with all parameters free to find the global minimum of the $-2\ln{\mathcal{L}}$, denoted as $-2\ln{\mathcal{L}}(\hat{\theta})$. Then, a grid is defined in $\theta_\nu$ and $\phi_\nu$, and for each point in the grid, a minimization is performed with $\theta_\nu$ and $\phi_\nu$ fixed at the corresponding values. This yields a $-2\ln{\mathcal{L}}$ value for each alternative hypothesis, which we denote $-2\ln{\mathcal{L}}(\theta')$. According to Wilks' theorem \cite{Wilks:1938dza} the $-2\Delta\ln{\mathcal{L}} = -2 [\ln{\mathcal{L}}(\theta_\text{true}) - \ln{\mathcal{L}}(\hat{\theta})]$ between the best fit and true parameter values follows a $\chi^2$-distribution with 2 degrees of freedom when profiling over the nuisance parameters. The $-2\Delta\ln{\mathcal{L}} = -2 [\ln{\mathcal{L}}(\theta') - \ln{\mathcal{L}}(\hat{\theta})]$ between the best fit and the alternative hypothesis can then be used to determine a confidence level for each point in the grid and draw uncertainty contours for the neutrino arrival direction. The reconstructed neutrino arrival directions, along with uncertainty contours derived from the profile likelihood scans for the three example events with one realization of noise each, are displayed in the top row of Figure \ref{fig_profile_llh}.

The validity of the estimated uncertainties can be verified by repeating the reconstruction many times using the same true parameter values and different realizations of noise. For any confidence level, we can then count how many times the true value would have been within the corresponding uncertainty contours, which gives the coverage. The confidence versus coverage is shown in the bottom row of Figure \ref{fig_profile_llh} for 100 trials of each of the three example events. The figure shows that the coverage is correct for any confidence level for the true parameter values tested here. Using the likelihood description presented in this work, we can thus reconstruct neutrino signals in radio detectors with band-limited noise and correctly estimate uncertainties on the reconstructed parameter values.

In order to investigate the impact of using the likelihood description, we repeat the reconstruction using the $\chi^2$ as expressed in Equation~\eqref{eq_chi2} as the objective function which does not take the correlations between bins into account. In this case, the reconstruction does successfully return best-fit parameters, however, two issues are present. First, if the $\chi^2$ is assumed to be $\chi^2$-distributed and used to estimate uncertainties on the reconstructed parameters, the coverage is completely wrong, which is shown in the dotted lines in the bottom row of Figure \ref{fig_profile_llh}. Hence, uncertainties can not be estimated for measured neutrinos, as resimulation with different noise realizations is only possible in simulation studies. Second, the reconstructed parameters for 100 trials have a significantly larger spread around the true values. The spread of the seven reconstructed parameters for 100 trials of the three example events using the two objective functions is shown in Table \ref{tab_llh_vs_chi2}. This result emphasizes that to achieve the best possible reconstruction resolution of signals in radio detectors, the correct likelihood description is needed, which employs the information about the correlations in the noise.

\begin{table}[!htb]
      \centering
        \begin{tabular}{|c|c||c|c|c|c|c|c|c|} 
            \hline
Event & Method & $\sigma_{E_\text{shower}}$ & $\sigma_{\theta_\nu}$ & $\sigma_{\phi_\nu}$ & $\sigma_{r_\text{vertex}}$ & $\sigma_{\theta_\text{vertex}}$ & $\sigma_{\phi_\text{vertex}}$ & $\sigma_{t_0}$ \\ \hline \hline
1
& \begin{tabular}[x]{@{}c@{}} $-2\ln{\mathcal{L}}$ \\  $\chi^2$\end{tabular}
& \begin{tabular}[x]{@{}c@{}} $28\%$ \\ $30\%$ \end{tabular}
& \begin{tabular}[x]{@{}c@{}} $1.4^{\circ}$ \\ $1.9^{\circ}$ \end{tabular}
& \begin{tabular}[x]{@{}c@{}} $9.6^{\circ}$ \\ $12^{\circ}$ \end{tabular}
& \begin{tabular}[x]{@{}c@{}} $134\,\text{m}$ \\ $154\,\text{m}$ \end{tabular}
& \begin{tabular}[x]{@{}c@{}} $0.59^{\circ}$ \\ $0.67^{\circ}$ \end{tabular}
& \begin{tabular}[x]{@{}c@{}} $0.099^{\circ}$ \\ $0.12^{\circ}$ \end{tabular}
& \begin{tabular}[x]{@{}c@{}} $0.15\,\text{ns}$ \\ $0.18\,\text{ns}$ \end{tabular}
\\ \hline
2
& \begin{tabular}[x]{@{}c@{}} $-2\ln{\mathcal{L}}$ \\  $\chi^2$\end{tabular}
& \begin{tabular}[x]{@{}c@{}} $40\%$ \\ $66\%$ \end{tabular}
& \begin{tabular}[x]{@{}c@{}} $4.3^{\circ}$ \\ $7.0^{\circ}$ \end{tabular}
& \begin{tabular}[x]{@{}c@{}} $2.2^{\circ}$ \\ $6.1^{\circ}$ \end{tabular}
& \begin{tabular}[x]{@{}c@{}} $31\,\text{m}$ \\ $42\,\text{m}$ \end{tabular}
& \begin{tabular}[x]{@{}c@{}} $0.044^{\circ}$ \\ $0.072^{\circ}$ \end{tabular}
& \begin{tabular}[x]{@{}c@{}} $0.11^{\circ}$ \\ $0.16^{\circ}$ \end{tabular}
& \begin{tabular}[x]{@{}c@{}} $0.16\,\text{ns}$ \\ $0.24\,\text{ns}$ \end{tabular}
\\ \hline
3
& \begin{tabular}[x]{@{}c@{}} $-2\ln{\mathcal{L}}$ \\  $\chi^2$\end{tabular}
& \begin{tabular}[x]{@{}c@{}} $7.5\%$ \\ $16\%$ \end{tabular}
& \begin{tabular}[x]{@{}c@{}} $1.4^{\circ}$ \\ $2.5^{\circ}$ \end{tabular}
& \begin{tabular}[x]{@{}c@{}} $1.7^{\circ}$ \\ $2.9^{\circ}$ \end{tabular}
& \begin{tabular}[x]{@{}c@{}} $32\,\text{m}$ \\ $66\,\text{m}$ \end{tabular}
& \begin{tabular}[x]{@{}c@{}} $0.20^{\circ}$ \\ $0.40^{\circ}$ \end{tabular}
& \begin{tabular}[x]{@{}c@{}} $0.031^{\circ}$ \\ $0.051^{\circ}$ \end{tabular}
& \begin{tabular}[x]{@{}c@{}} $0.055\,\text{ns}$ \\ $0.11\,\text{ns}$ \end{tabular}
            \\ \hline
        \end{tabular}
    \caption{Spread (standard deviation) of the reconstructed parameter values for 100 trials of the three example events listed in Table \ref{tab_events} minimizing the $-2\ln{\mathcal{L}}$ presented in Equation~\eqref{eq_multivariate_normal} in this work compared to minimizing the $\chi^2$ expressed in Equation~\eqref{eq_chi2}, which does not take correlation into account.}
     \label{tab_llh_vs_chi2}
\end{table}

\section{Summary and outlook}

In this work, we have presented a likelihood description of a deterministic signal in a radio detector with band-limited noise. The likelihood allows us to obtain uncertainties on the reconstructed parameters, which was not possible with previous approaches. We have demonstrated with a toy-model study using this likelihood description that the uncertainties have correct coverage. We also find that the Likelihood yields smaller uncertainties for in-ice neutrino reconstruction compared to the previously employed $\chi^2$ function that does not take into account correlations. 

The likelihood description is generally a strong tool with several additional applications currently being investigated. Since the likelihood ratio is the best test statistic to compare two hypotheses, it can be used to separate signal from background. This fact can potentially be used to improve the identification of cosmic-ray signals in in-ice radio detectors, like in the cosmic-ray search recently performed with the ARIANNA detector \cite{Arianna:2021lnr}. Second, the likelihood description provides a way to calculate a goodness-of-fit for a reconstructed signal. For a reconstructed signal that describes the measured trace well, the $-2\ln{\mathcal{L}}$ in Equation~\eqref{eq_multivariate_normal} (ignoring the constants) should be $\chi^2$-distributed with degrees of freedom equal to the number of samples in the trace from which a p-value can be calculated. Finally, the covariance matrix discussed in Section \ref{sec_noise_model} can be used to calculate the Fisher information matrix for a detector and set of signal parameters. The inverse Fisher information matrix gives an estimate of the uncertainties and correlations of the signal parameters without the need to perform any reconstruction. The likelihood description presented in this work is thus potentially also useful in the prospect of end-to-end detector optimization.

\bibliographystyle{JHEP}
\bibliography{refs}

\end{document}